\documentclass{aa}
\usepackage{graphicx}
\usepackage{amsmath}
\usepackage{amssymb}
\usepackage{txfonts}
\usepackage{xcolor}

\newcommand{\rosl}{{\it Roentgen Satellit}}
\newcommand{\ros}{{\it ROSAT}}

\newcommand{\xmm}{{\it XMM-Newton}}
\newcommand{\xmml}{{\it X-ray Multi-Mirror Mission}}

\newcommand{\eros}{{eROSITA}}
\newcommand{\erosl}{{extended Roentgen Survey with an Imaging Telescope Array}}
\newcommand{\nicer}{{NICER}}
\newcommand{\nicerl}{{Neutron star Interior Composition Explorer}}

\newcommand{\swift}{{\it Neil Gehrels Swift Observatory}}

\newcommand{\eso}{ESO-VLT}

\newcommand{\saltl}{{Southern African Large Telescope}}

\newcommand{\nh}{N_{\rm H}}

\newcommand{\forstl}{{Focal Reducer/low dispersion Spectrograph 2}}
\newcommand{\forst}{{FORS2}}

\def \magoe{\object{RX~J1856.5--3754}}
\def \magzs{\object{RX~J0720.4--3125}}

\def \magot{\object{RBS~1223}}

\def \magzf{\object{RX~J0420.0--5022}}

\def \otos{\object{J1317}}

\def \osts{\object{PSR~J0250+5854}}
\def \gleam{\object{GLEAM-X J162759.5--523504.3}}
\def \onoo{\object{PSR~J0901--4046}}
\def \onot{\object{PSR~J1903+0433g}}

\def \jotos{\object{eRASSU~J131716.9--402647}}

\def \fluxcgs{erg~s$^{-1}$~cm$^{-2}$}

\begin{document} 
%

\title{Detection of pulsed X-ray emission from the isolated neutron star candidate eRASSU~J131716.9--402647\thanks{Based on observations obtained with \xmm, an ESA science mission with instruments and contributions directly funded by ESA Member States and NASA (observation 0921280101)}}

\author{J.~Kurpas\inst{1,2}
   \and A.~D.~Schwope\inst{1}
   \and A.~M.~Pires\inst{1,3}
   \and F.~Haberl\inst{4}
}
\institute{Leibniz-Institut f\"ur Astrophysik Potsdam (AIP), An der Sternwarte 16, 14482 Potsdam, Germany
   \email{jkurpas@aip.de} 
   \and
   Potsdam University, Institute for Physics and Astronomy, Karl-Liebknecht-Stra\ss e 24/25, 14476 Potsdam, Germany
   \and
   Center for Lunar and Planetary Sciences, Institute of Geochemistry, Chinese Academy of Sciences, 99 West Lincheng Rd., 550051 Guiyang, China
   \and
   Max-Planck-Institut f\"ur extraterrestrische Physik, Giessenbachstra\ss e 1, 85748 Garching, Germany
}
\date{Received ...; accepted ...}
\titlerunning{Pulsed X-ray emission from \otos}
\authorrunning{J.~Kurpas et al.}
\abstract
{The X-ray source \jotos\ was recently identified from observations with {Spectrum Roentgen Gamma} (SRG)/eROSITA as a promising X-ray dim isolated neutron star (XDINS) candidate on the premise of a soft energy distribution, absence of catalogued counterparts, and a high X-ray-to-optical flux ratio. Here, we report the results of a multi-wavelength observational campaign with \xmm, \nicer\ and the FORS2 instrument at the \eso. We found in both the \xmm\ and \nicer\ data that the X-ray emission is strongly pulsed at a period of $12.757$~s (pulsed fraction $p_\mathrm{f} = (29.1 \pm 2.6)$\% in the 0.2--2~keV band). The pulse profile is double-humped, and the pulsed fraction increases with energy. The \xmm\ and \nicer\ epochs allow us to derive a 3$\sigma$ upper limit of $\dot{P}\leq 8\times 10^{-11}$~s~s$^{-1}$ on the spin-down rate of the neutron star. The source spectrum is well described by a purely thermal continuum, either a blackbody with $kT\sim95$~eV or a magnetised neutron star atmosphere model with $kT \sim 35$~eV. Similarly to other thermally emitting isolated neutron stars, we found in either case strong deviations from the continuum, a broad absorption feature at energy $\sim260$~eV and a narrow one around $590$~eV. The FORS2 instrument at \eso\ has not detected the optical counterpart ($m_\mathrm{R}>27.5$~mag, $5\sigma$ detection limit), implying an X-ray-to-optical flux ratio of $10^4$ at least. The properties of \jotos\ strongly resemble those of a highly magnetised isolated neutron star and favour an XDINS or high-B pulsar nature.}
\maketitle
\section{Introduction\label{sec_intro}}

The seven known X-ray dim isolated neutron stars (XDINSs) are famous for their thermal spectra, which are seemingly devoid of accretion or magnetospheric activity, have long spin periods ranging from 3~s to 17~s, and have higher magnetic field strengths ($10^{13}$--$10^{14}$~G) than rotation-powered pulsars of similar ages \citep[][]{2009ASSL..357..141T}. Interestingly, a more complex surface temperature distribution and high energy excess has been detected in their cumulative X-ray spectra only recently \citep{2017PASJ...69...50Y, 2019PASJ...71...17Y, 2020ApJ...904...42D, 2022MNRAS.516.4932D}. Although the currently observed sample is small, these objects might constitute a substantial part of the Galactic population of isolated neutron stars (INS). This is implied from their estimated birthrate, which could be on par with that of the much more numerous class of rotation-powered pulsars \citep{2008MNRAS.391.2009K}.

The seven confirmed XDINSs have all been identified in \rosl\ (\ros) observations \citep[][]{2007Ap&SS.308..181H}. Additional searches with \ros\ and the \swift, the ever-growing dataset from the pointed \xmml\ (\xmm), and the new all-sky survey at X-ray energies conducted with the \erosl\ (\eros) instrument on board the {Spectrum Roentgen Gamma} (SRG) mission \citep{2021A&A...647A...1P}, have led to the identification of new candidates \citep[][]{2008ApJ...672.1137R, 2009A&A...504..185P, 2022MNRAS.509.1217R, 2022A&A...666A.148P, 2023A&A...674A.155K}.

One of the recently identified candidates is the X-ray source \jotos\ \citep[hereafter dubbed \otos;][]{2023A&A...674A.155K}. Detected in the \eros\ all-sky surveys, with a predominantly thermal X-ray spectrum that is best fit by a single absorbed blackbody model with an effective temperature of $110$~eV and a high X-ray-to-optical flux ratio that implies a compact nature, it was found to fit well into the known XDINS population. However, the data available at that time did not allow ruling out the existence of non-thermal emission. Similarly, the low photon count and survey exposure gaps\footnote{\eros\ covered the source position five times between January 2020 and January 2022; see \citet{2023A&A...674A.155K} for details.} prevent a meaningful timing analysis. Therefore, an unambiguous classification of the source has not been possible yet, as is the case for other proposed XDINS candidates. To shed light on its nature, we carried out follow-up observations of \otos\ with the \nicerl\ \citep[\nicer;][]{2016SPIE.9905E..1HG}, \xmm\ \citep{2001A&A...365L...1J}, and the \forstl\ (\forst) instrument at the ESO Very Large Telescope \citep[\eso;][]{1998Msngr..94....1A}.\\
\indent In this paper, we report the results of this observational campaign and discuss the implications for the nature of the source. We start in Sect.~\ref{sec_obs} by describing the new data sets and reduction steps. The results of the X-ray spectral and timing analysis and optical imaging are presented in Sect.~\ref{sec_analysis}. Our conclusion and outlook are outlined in Sect.~\ref{sec_disc}.

\section{Data reduction\label{sec_obs}}

\begin{table}
\caption{List of all X-ray and optical observations.
\label{tab_obsinfo}}
\centering

\begin{tabular}{llrrrrrrr}
\hline\hline
OBSID/ & MJD\,\tablefootmark{(a)} & $T_\mathrm{exp}$\,\tablefootmark{(b)} & Counts\,\tablefootmark{(c)}\\
Programme ID & & [s] & \\
\hline
\multicolumn{4}{l}{\nicer}\\
\hline
6572020101 & 60032.590349 &   6338 &  6620 \\
6572020102 & 60032.975650 &  18789 & 19293 \\
6572020103 & 60034.008011 &  20067 & 25648 \\
6572020104 & 60035.043497 &   3568 &  3814 \\
\hline
\multicolumn{4}{l}{\xmm}\\
\hline
0921280101 & 60133.583028 & 36355 & 9312\\
\hline
\multicolumn{4}{l}{\eso}\\
\hline
111.259R.001 & 60077.223172 & 3088 & \\
111.259R.001 & 60082.114395 & 3088 & \\
\hline
\end{tabular}
\tablefoot{For \xmm, we provide the EPIC-pn values in the table.
\tablefoottext{a}{Modified Julian date relative to the start of the observation.}
\tablefoottext{b}{Where applicable, the exposure times are free from periods of high background activity, but they are uncorrected for vignetting and CCD dead time.}
\tablefoottext{c}{Total X-ray photon events detected in the 0.3--2.0~keV energy band.}
}
\end{table}

\subsection{X-ray observations}

The X-ray source \otos\ was observed by \nicer\ for $49$~ks in March and April 2023 (Table~\ref{tab_obsinfo}). The raw data files were reduced with the NICERDAS software tools (Version: 2023-08-22\_V011a) that are distributed with the HEASoft release 6.32.1. We extracted the event lists using the \texttt{nicerl2} task and adopted a conservative event screening by setting the "saafilt" variable to "yes" and "saafilt\_nicer" to "no", the "underonly\_range" was set from 0--50, the "overonly\_range" from 0--5, "cor\_range" to "1.5--$\star$" and "elv" to a value of 30. Verifying that no periods of high background remained in the single observation files, we merged all observations with \texttt{niobsmerge}. We extracted the spectrum with the \texttt{nicerl3-spect} pipeline and used the SCORPEON\footnote{\url{https://heasarc.gsfc.nasa.gov/docs/nicer/analysis_threads/scorpeon-overview/}} model for the background estimation. Similarly, the \texttt{nicerl3-lc} pipeline was used to extract light curves. We applied the barycentric correction to the 55375 cleaned and filtered events (0.3--2~keV band) with the \texttt{barycorr} task (ephemeris: JPLEPH.405, using the improved \xmm\ based sky position).

\begin{figure}[t]
\begin{center}
\includegraphics[width=\linewidth]{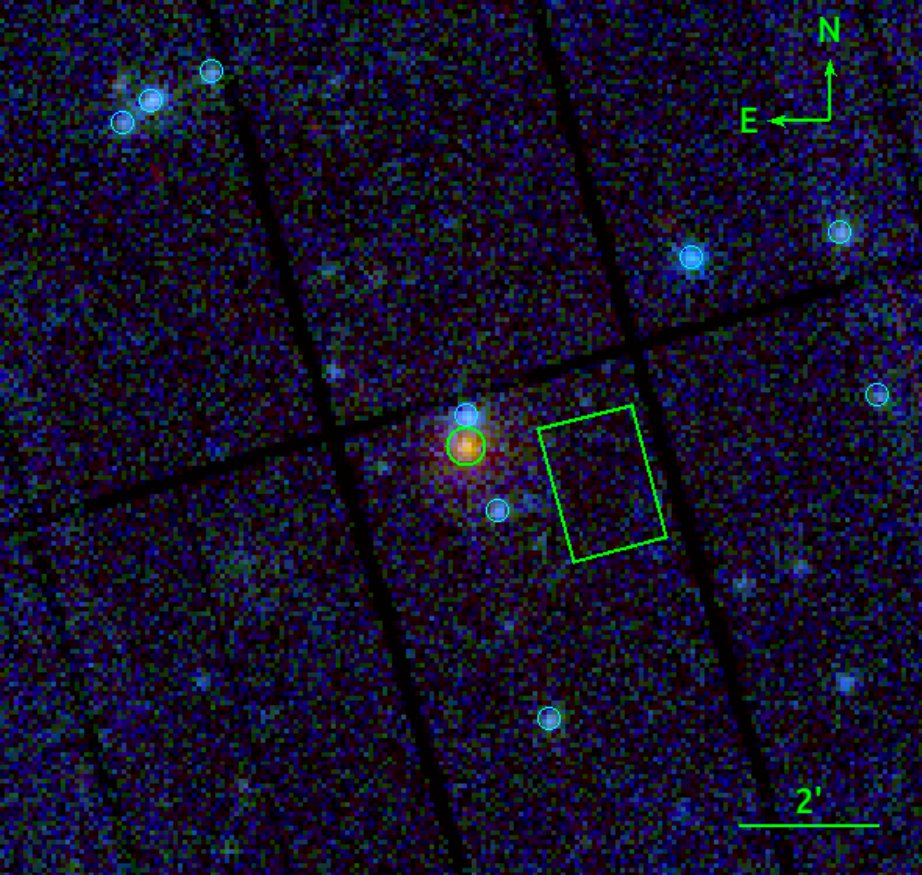}\vskip1pt
\end{center}
\caption{RGB \xmm\ EPIC image. Red indicates counts in the 0.2--1~keV band, green in the 1--2~keV band, and blue in the 2--12~keV band. The source and background regions used in the EPIC-pn analysis are marked by the green circle and box. The X-ray sources we used to correct the X-ray sky position of \otos\ are marked by cyan circles.}
\label{fig_epic_img}
\end{figure}

The \xmm\ observation of \otos\ was performed in July 2023 for $36$~ks (Table~\ref{tab_obsinfo}). The European Photon Imaging Camera's (EPIC) pn and MOS detectors, equipped with the THIN filter, were operated in full frame and small window mode (73.4~ms and 300~ms time resolution\footnote{\url{https://xmm-tools.cosmos.esa.int/external/xmm_user_support/documentation/uhb/epicmode.html}}, respectively). We extracted data products using the Science Analysis System (SAS; Version: 21.0.0) and found the observation to be free from periods of enhanced background activity. To determine the source position, we performed source detection with the task \texttt{edetect\_stack} in the standard \xmm\ energy bands 0.5--1~keV, 1--2~keV and 2--4.5~keV, using the exposures from both the EPIC-MOS and pn detectors. We then applied the task \texttt{eposcorr} to refine the astrometry. We compared the position of all detected X-ray sources with a detection likelihood value "EP\_DET\_ML" above 150 with those of objects in the Guide Star Catalog \citep[Version, 2.4.2;][]{2008AJ....136..735L}, whose positional errors are below 1.5\arcsec. The number of considered matches, the applied offset in RA and DEC from the initial \xmm\ source detection localisation, and the updated source position are shown in Table~\ref{tab_posphotpar}. The results of the source detection were used to define a background region on the same CCD chip as the target devoid of other X-ray sources. The optimal centring and extraction region for the target was determined with the task \texttt{eregionanalyse} in the 0.2--2.0~keV band. We decided to truncate the optimal size of 40\arcsec\ to a value of 16\arcsec\ to minimise contamination by a neighbouring hard X-ray source located 26\arcsec\ north of the target (Fig.~\ref{fig_epic_img}). The event lists were further cleaned by applying the recommended filters "$FLAG=0$", "$PATTERN\leq4$" for EPIC-pn and "$PATTERN \leq 12$" for EPIC-MOS. Likewise, we followed the \xmm\ data analysis threads\footnote{\url{https://www.cosmos.esa.int/web/xmm-newton/sas-threads}} to extract the spectra. For the timing analysis, the 10~374 source events from EPIC-pn (0.2--2.0~keV) were barycentric corrected with the \texttt{barycen} task (ephemeris: DE-405) using the updated source position.

\begin{table}
\caption{Positional and photometric parameters.
\label{tab_posphotpar}}
\centering

\begin{tabular}{lr|lrrrrrr}
\hline\hline
RA [\degr]            & 199.32096 & Mag. ZP                                  & 28.18\\
DEC [\degr]           & -40.44636 & $\sigma_\mathrm{sky}$                    & 0.08\\
POSERR [\arcsec]      & 0.23      & FWHM [px]                                & 3.41\\
b [\degr]             & 22.16     & FWHM ["]                                 & 0.86\\
Ref. sources\,\tablefootmark{(a)}          & 31        & $5\sigma$ detection limit                & 27.5\\
RA\_offset [\arcsec]  & 1.15      & absorbed $f_\mathrm{X}/f_\mathrm{opt}$   & 10801\\
DEC\_offset [\arcsec] & -0.24     &  &  \\
\hline
\end{tabular}
\tablefoot{.
\tablefoottext{a}{Number of matches considered by \texttt{eposcorr} to improve the X-ray sky position.}
}
\end{table}

\subsection{Optical observations}
Follow-up \eso\ observations of the field of \otos\ were carried out on May 12 and 18, 2023 (Table~\ref{tab_obsinfo}). We used the FORS2 instrument equipped with the \textit{R\_SPECIAL} filter to obtain deep imaging under generally photometric conditions. Three exposures affected by worsening seeing conditions ($>1.3\arcsec$) on May 12 were not included in the analysis. We applied the standard FORS2 reduction pipeline within the ESO Reflex environment \citep{2013A&A...559A..96F} to correct the images for bias and flat field. The individual science exposures were astrometrically corrected with astrometry.net \citep{2010AJ....139.1782L}. To stack the images, they were aligned with the task \texttt{wcsalign} \citep{astropy:2013, astropy:2018, astropy:2022}.

\section{Results\label{sec_analysis}}
\subsection{Timing analysis\label{sec_timing}}
\begin{figure}[t]
\begin{center}
\includegraphics[width=\linewidth]{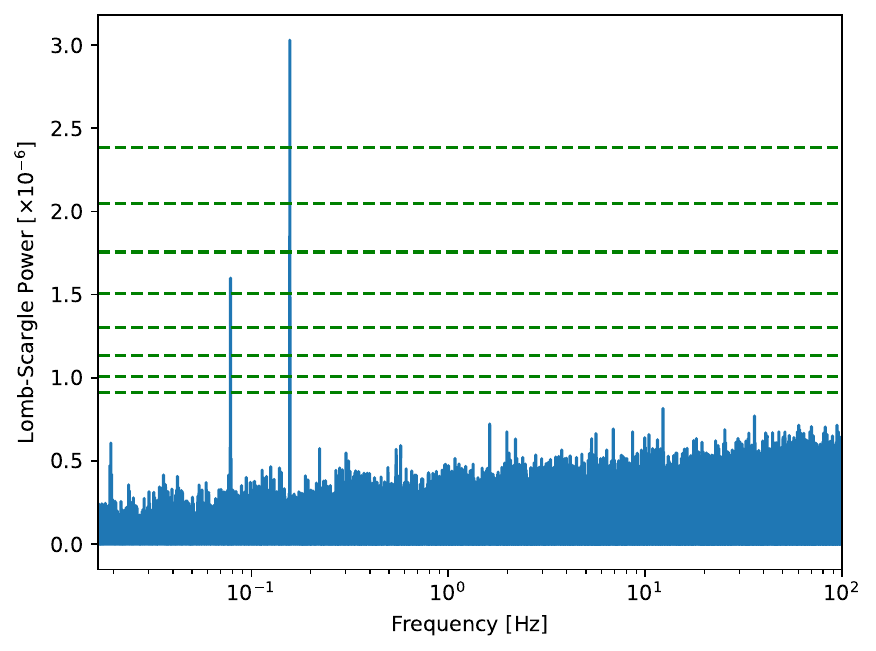}\vskip1pt
\end{center}
\caption{\nicer\ Lomb-Scargle periodogram. The green lines indicate the Lomb-Scargle power, which equals a significance of 1--8$\sigma$ (from bottom to top).}
\label{fig_l_s_period}
\end{figure}

In comparison to the science modes of \xmm\ we used, the time resolution of \nicer\ allows us to search a broader frequency range for periodic modulations. We binned the barycentre-corrected \nicer\ photon event list into a 1~ms light curve and applied the astropy \texttt{LombScargle} tool \citep{astropy:2013, astropy:2018, astropy:2022} to compute the Lomb-Scargle periodogram \citep{1976Ap&SS..39..447L,1982ApJ...263..835S} for spin periods between 0.01~s--60~s (Fig.~\ref{fig_l_s_period}). The key parameters of the search we conducted are summarised in Table~\ref{tab_timepar}. The periodogram shows a significant periodic signal and its first harmonic at a period of $12.76$~s and $6.37$~s, respectively. The existence of this modulation is confirmed by the \xmm\ data because the Lomb-Scargle periodogram computed in the range of 0.3--60~s, using photons from EPIC-pn binned into a 150~ms light curve, also detects the modulation with high significance. \xmm\ allows us to verify that the nearby but fainter X-ray emitting source north of the target position (see image in Fig.~\ref{fig_epic_img}) is not the origin of the pulsed emission. A corresponding search over the \xmm\ light curve of the source did not reveal any significant periodic signal because the highest peak possesses a false-alarm probability very close to 1.

To improve the period estimation and derive its significance, we applied the Bayesian-based folding method of \cite{1996ApJ...473.1059G}. The method calculates a prescription of frequency-dependent odds ratio $O_\mathrm{m}$ that the data favour a periodic model with $m$ phase bins over the null (unpulsed) model. Following \citet[][]{2000ApJ...540L..25Z}, we adopted a characteristic number of $m_\mathrm{max} = 12$ and a frequency interval of 20~$\mu$Hz centred on the fundamental frequency $\nu_0 =0.078388$~Hz. The resulting odds ratio values of $O_\mathrm{per}^\mathrm{\star}= 9 \times 10^{29}$ and $5 \times 10^{48}$, obtained for \nicer\ and \xmm, respectively, give a probability of virtually 100\% that a periodic signal is present in the tested frequency range (see Table~\ref{tab_timepar} for details). The corresponding 68\% confidence intervals are $P_\nicer = 12.757129(16)$~s and $P_{XMM} = 12.75707(7)$~s. The accuracy on the period is not high enough to phase-connect the two data sets because the more accurate \nicer\ timing solution predicts $683991.2 \pm 0.9$ spin cycles between the time of phase zero of both observations. Nonetheless, the current estimates rule out spin-down values in excess of $\dot{P} \leq 8\times 10^{-11}$~s~s$^{-1}$, at the $3\sigma$ confidence level.

\begin{figure}[t]
\begin{center}
\includegraphics[width=\linewidth]{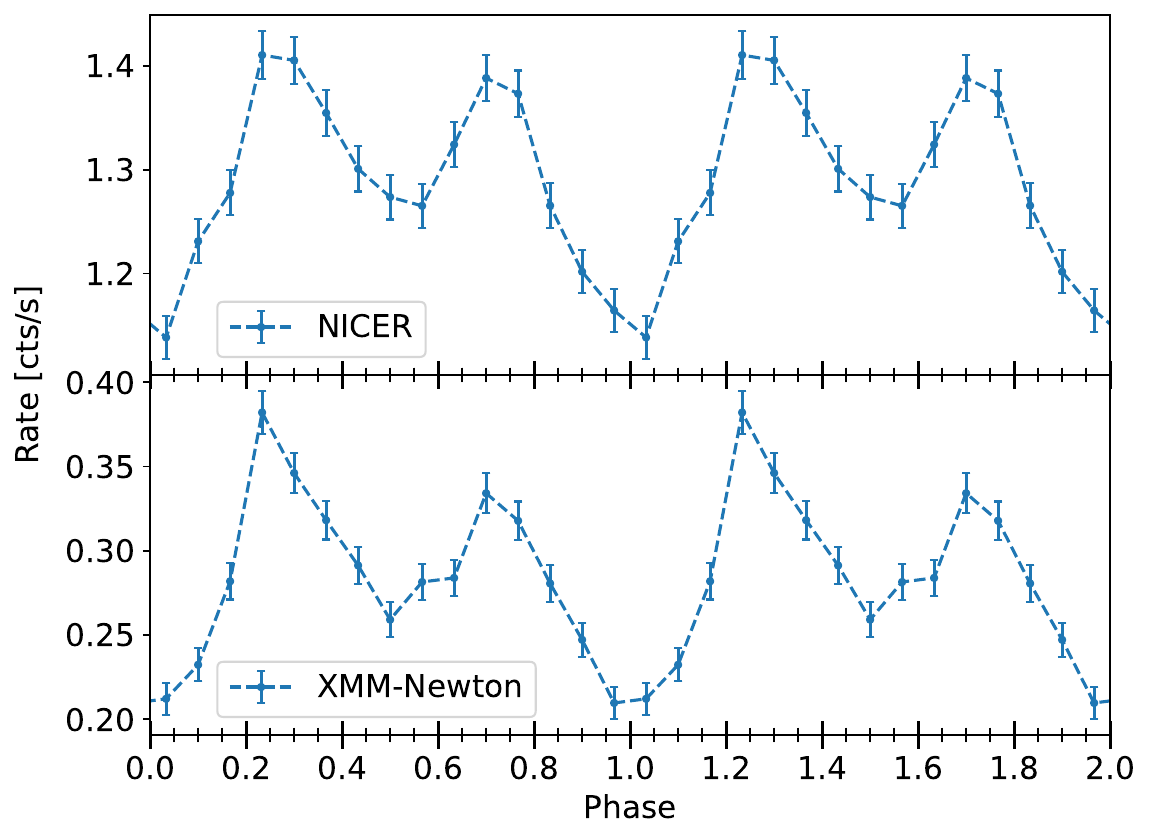}\vskip1pt
\end{center}
\caption{Phase-folded light curves showing the pulse profile of \otos\ in the 0.3--2~keV (\nicer) and 0.2--2~keV (\xmm) energy band. Two cycles are shown for visual purposes.}
\label{fig_p_profile}
\end{figure}

\begin{figure}[t]
\begin{center}
\includegraphics[width=\linewidth]{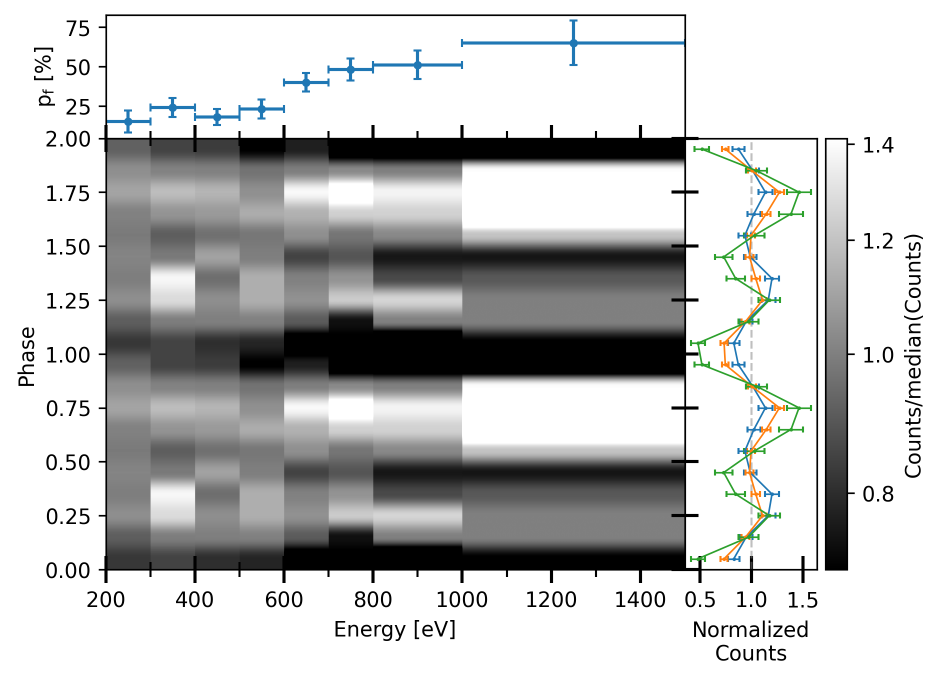}\vskip1pt
\end{center}
\caption{Dependence of the pulse profile as a function of energy and phase. The number of detected photons in EPIC-pn, normalised by the median value derived from the sum of all photons that were detected in a certain energy bin column, is colour-coded. At the top, we indicate the pulsed fraction, determined from events in a certain energy bin, whereas to the right, we plot the normalised light curves in the energy bands 0.2--0.4~keV (blue), 0.4--0.8~keV (orange), and 0.8--2~keV (green). They were computed by summing up all events in a certain phase bin and energy range and normalising again by the median number of counts in the given energy range and all phase bins.}
\label{fig_p_e_dependence}
\end{figure}

With the corresponding best period estimates, we folded both light curves in phase and chose the minimum of the pulse profile as the time of phase zero (see Table~\ref{tab_timepar} and Fig.~\ref{fig_p_profile}). Consistent with the results of the Lomb-Scargle the pulse profile is clearly double-humped. Interestingly, the minima between the pulses are of different depth and the pulses are of different height, the second being weaker than the first. We derived pulsed fractions of $(10.7 \pm 1.2)$\% (0.3--2.0~keV) for \nicer\ and $(29.1 \pm 2.6)$\% (0.2--2.0~keV) for \xmm, respectively. The background was not subtracted for either value. Thus, the discrepancy may be explained by a stronger background event contamination for \nicer. In Fig.~\ref{fig_p_e_dependence}, we show the energy and phase dependence of the pulse for EPIC-pn. The pulse at phase 0.7 seems to be strongest at intermediate to high energies, whereas the pulse around phase 0.25 seems to be strongest at soft energies. Similarly, the pulsed fraction increases with energy from 15(7)\% (0.2--0.3~keV) to 65(14)\% (1--1.5~keV).

\begin{table}
\caption{Timing parameters from the \nicer\ and \xmm\ observation.
\label{tab_timepar}}
\centering

\begin{tabular}{llll|rrrrr}
\hline\hline
 & \nicer & \xmm \\
\hline
$N_\mathrm{photon}$ & 55 375 & 10 374\\
Energy range [keV] & 0.3 -- 2.0 & 0.2 -- 2.0\\
$\Delta T$\,\tablefootmark{(a)} [s] & 246 057 & 36 349 \\
$\nu_\mathrm{LS, min}$ -- $\nu_\mathrm{LS, max}$ [Hz] & 0.0167 -- 100 & 0.0167 -- 3.33\\
LS significance\,\tablefootmark{(b)} & 5.4$\sigma$ & 6.7$\sigma$\\
$\Delta \nu_\mathrm{GL}$ [$\mu$Hz] & 20 & 20 \\
O$_\mathrm{per}^\mathrm{\star}$ & $ 5 \times 10^{31}$ & $5 \times 10^{48}$ \\
$p_{\nu_1, \nu_2}$ [\%]\,\tablefootmark{(c)}& $\approx100$ & $\approx100$\\
Frequency [Hz] & 0.07838754(10) & 0.0783879(5)\\
Period [s]   & 12.757129(16)  & 12.75707(7)\\
Pulsed-fraction\,\tablefootmark{(d)} [\%] & $10.7 \pm 1.2$ & $29.1 \pm 2.6$\\
Time-of-phase zero\,\tablefootmark{(e)} & 60032.590341 & 60133.582978\\
\hline
\end{tabular}
\tablefoot{
\tablefoottext{a}{Time between first and last photon}
\tablefoottext{b}{For the fundamental peak, we show the significance that the observed Lomb-Scargle power is not due to white noise. The chance that the observed peak is due to noise was computed from the false-alarm probability \citep{2008MNRAS.385.1279B}.}
\tablefoottext{c}{Probability of detection of a periodic signal in the frequency interval we probed. Computed via $\mathrm{O}_\mathrm{per}^{\star}/(1+\mathrm{O}_\mathrm{per}^{\star})$, with $\mathrm{O}_\mathrm{per}^{\star} = \left(\ln\frac{\nu_2}{\nu_1}\right)^{-1} \int^{\nu_1}_{\nu_2} \frac{d\nu}{\nu} O_\mathrm{per}(\nu)$ \citep[see][for more details]{2000ApJ...540L..25Z}.}
\tablefoottext{d}{Computed from the maximum and minimum light-curve count-rate values ($R$) via $p_\mathrm{f} = \frac{R_\mathrm{max}-R_\mathrm{min}}{R_\mathrm{max}+R_\mathrm{min}}$ (Fig.~\ref{fig_p_profile}).}
\tablefoottext{e}{Reference time in MJD used to align the \nicer\ and \xmm\ pulse profiles.}
}
\end{table}


\subsection{X-ray spectral analysis}

We used \texttt{XSPEC} \citep[version 12.13.0,][]{1996ASPC..101...17A} to fit spectral models to the \xmm\ and \nicer\ data sets. The spectra were binned with at least 25 counts per spectral bin to allow for the use of the $\chi^2$ statistic. We accounted for the interstellar absorption by applying the \texttt{tbabs} model with elemental abundances described in \cite{2000ApJ...542..914W}. When spectra from several instruments (EPIC-pn, EPIC-MOS1, EPIC-MOS2, and \nicer) were fitted simultaneously, we added a constant factor to account for calibration uncertainties. When fitting the \nicer\ spectrum, the background was simultaneously fitted adopting the SCORPEON model. We found that all spectra are affected by the hard source located north of the target. To account for the level of contamination and estimate its dependence with energy, we extracted the spectrum of the hard source from the EPIC-pn exposure. We used a small aperture of 6\arcsec\ to prevent bias from the brighter point spread function (PSF) of the target. Due to the low count number (357 events extracted in the 0.2--12~keV energy range), we adopted the Bayesian X-ray Analysis \citep[BXA;][]{2014A&A...564A.125B} package to fit the unbinned data. The spectrum of the contaminant is well described by an absorbed power-law model with a photon index of $\Gamma = 1.81^{+0.15}_{-0.13}$, $\nh = (10 \pm 4) \times 10^{20}$~cm$^{-2}$ and $f_\mathrm{X} =8.4(5) \times 10^{-14}$~\fluxcgs (0.2--10~keV), typical of that of an active galactic nucleus (AGN). The column density is consistent with the Galactic value of $6.7\times 10^{20}$~cm$^{-2}$ in the direction of the source \citep{2016A&A...594A.116H}. We added this power-law model as an additional background component to fits of the \nicer\ spectrum. For \xmm\ the contamination can be mitigated if the energy range is constrained to 0.2--2.0~keV. We found \xmm\ and \nicer\ to give consistent results when \nicer\ is limited to the energy range of 0.4--10~keV. We list the fit results of all models we used in Table~\ref{tab_fitinfo}.

\begin{figure}[t]
\begin{center}
\includegraphics[width=\linewidth]{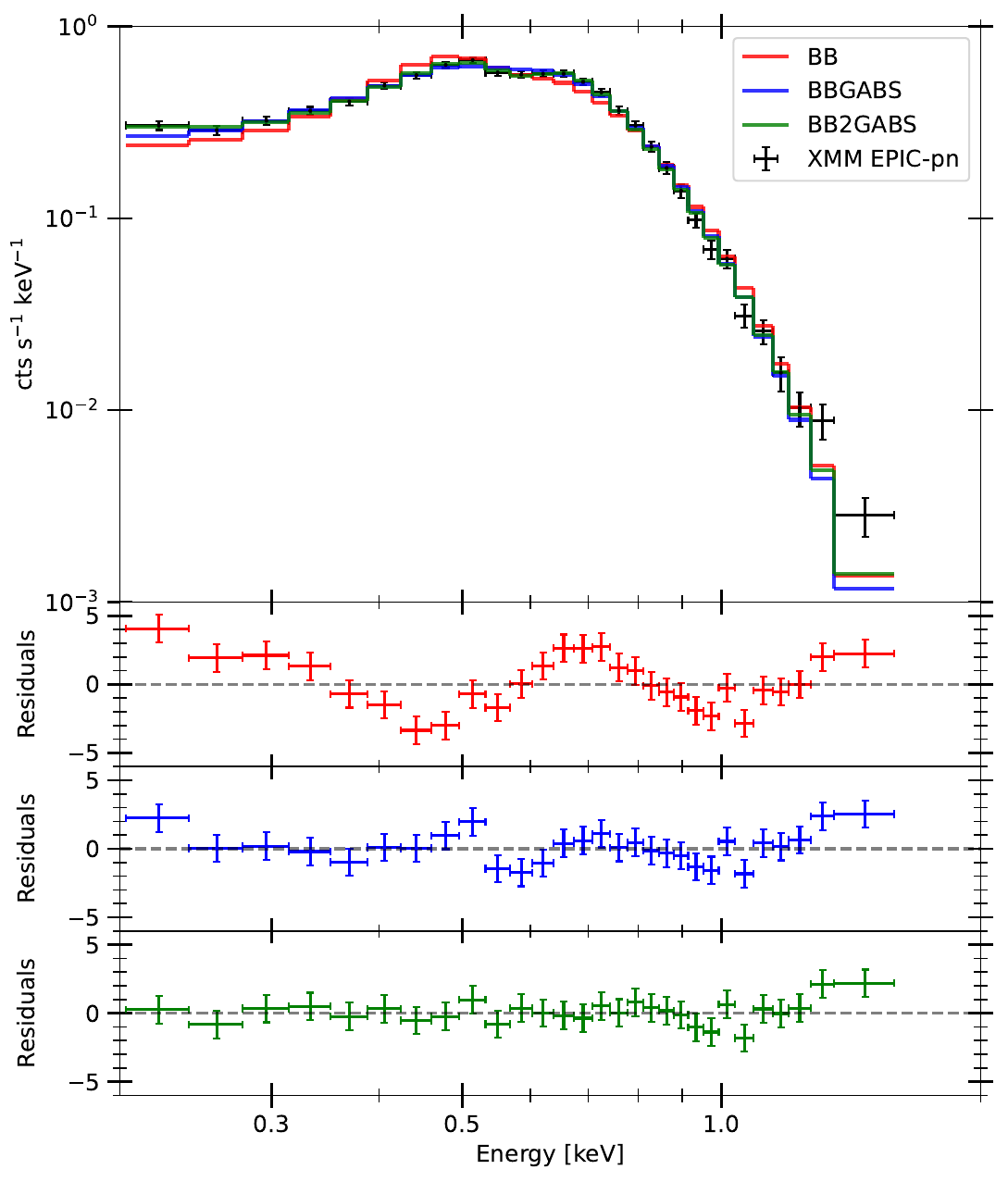}\vskip1pt
\end{center}
\caption{EPIC-pn spectrum in the energy range 0.2--2~keV. We indicate the best model and residuals for a single absorbed \texttt{BB} model (red), an absorbed \texttt{BB} with a single absorption component (blue), and an absorbed \texttt{BB} with two absorption components (green).}
\label{spec_plt_bb}
\end{figure}

We began the spectral analysis by fitting a single blackbody (\texttt{BB}), neutron star atmosphere \citep[\texttt{NSA};][]{1995ASIC..450...71P, 1996A&A...315..141Z} or power-law (\texttt{PL}) model to the EPIC-pn spectrum. This approach is motivated by the fact that the spectral analysis of the \eros\ data could not exclude any of these models \citep{2023A&A...674A.155K}. For all three models, we found high systematic residuals, a poor fit quality, and column densities in excess of the Galactic value (see Table~\ref{tab_fitinfo} and Fig.~\ref{spec_plt_bb}, for the \texttt{BB} case). The addition of a \texttt{PL} component on the thermal models (\texttt{BB}, \texttt{NSA}), as usually observed as a hard tail in the spectrum of middle-aged spin-powered pulsars \citep[e.g.][]{2022A&A...661A..41S}, does not improve the fit. Combining multiple \texttt{BB} components, which is indicative of a more complex surface temperature distribution, did not improve the fit either. Interestingly, we found that the addition of a multiplicative broad Gaussian absorption line (\texttt{GABS} model) at the low-energy end of the spectrum ($200-400$~eV) significantly improved the fit quality. For a \texttt{PL} continuum (combined with a single \texttt{GABS} component), the resulting fit shows a good agreement between data and model and no significant residuals that might imply the necessity for additional model components. The fit of the thermal models (both \texttt{BB} and \texttt{NSA}) with one \texttt{GABS} component, although an improvement with respect to the single-component case, still shows high $\chi^2_\nu$ values and systematic residuals consistently around energies of $550-600$~eV (see Fig.~\ref{spec_plt_bb}, again for the \texttt{BB} case). The addition of a second \texttt{GABS} component improves the results in both cases, but it introduces degeneracy between $\nh$ and the parameters of the lines. Fixing the widths of the two Gaussian components to their best-fit values, $\sigma_1 = 150$~eV and $\sigma_2 = 15$~eV, leads to slightly more accurate line energies and smaller errors on the $\nh$ (shown for the \texttt{BB} case in Table~\ref{tab_fitinfo}), but does not allow us to constrain the exact value of the column density. Thus, the $\nh$ values we obtained can only be regarded as an upper limit, but they strongly favour a Galactic source.

For a BB continuum, we studied the significance of the detected lines by computing the false-negative rates (an existing feature is not identified) and false-positive rates (the fit statistics are improved only by chance when absorption components are added). We used the best-fit \texttt{BB2GABS} model and the XSPEC \texttt{fakeit} command to simulate 1000 spectra and fitted them with a single \texttt{BB}, \texttt{BBGABS}, and \texttt{BB2GABS} model. We found that a \texttt{BB} model was never preferred over the \texttt{BBGABS} fit and that a \texttt{BBGABS} only gave better fits than a \texttt{BB2GABS} model in 2\% of the cases. Since the line parameters were generally well recovered, the simulations imply very low false-negative rates close to 0\% for the broad feature at $\sim 260$~eV and 2\% for the narrow line at $\sim 590$~eV. For the false-positive case, we simulated 1000 \texttt{BB} and \texttt{BBGABS} spectra each and then conducted fits with the simulated model and with one model that contained an additional absorption component. In both cases, we found that similarly large improvements in the fit statistic ($\Delta \chi^2_\nu =2.4$ for the broad feature and $\Delta \chi^2_\nu =0.65$ for the narrow one) by the addition of a line were never observed in the simulations. This implies false-positive rates close to 0\% for both features. We conclude that both features seem to be detected with very high significance.

We next tried fits with other absorption feature components available in XSPEC (\texttt{GAUSS} and \texttt{edge}), combining them with a \texttt{BB} continuum. Fits with the \texttt{GAUSS} component reproduce the results from the \texttt{GABS} models. On the other hand, three \texttt{edge} components are necessary to properly fit the spectrum ($\chi^2_\nu(\nu) = 1.18(20)$), suggesting a possible third line at $864^{+22}_{-25}$~eV. However, we note that the resulting $\nh = 14^{+4}_{-3}\times 10^{20}$~cm$^{-2}$ significantly exceeds the Galactic value and that the fit statistics is worse than in the \texttt{GABS} and \texttt{GAUSS} cases.

The \texttt{NSA} model fits have shown that in comparison to non-magnetised models, better $\chi^2_\nu$ values are obtained when the magnetic field is taken into account. They allow us to obtain distance estimates. We do not list them in Table~\ref{tab_fitinfo}, but found the largest distances for magnetised \texttt{NSA} models with two absorption lines, converging to $100^{+70}_{-16}$~pc ($10^{12}$~G) and $130^{+160}_{-18}$~pc ($10^{13}$~G), respectively. These may be comparable (within 2--3$\sigma$) to those inferred for the known population of XDINSs \citep[see e.g.][and references therein]{2009A&A...497..423M}, but could be too small considering the much lower flux of \otos.

Alternatively, composite models consisting of two blackbody components or of a blackbody plus power-law, in both cases modified by a Gaussian absorption feature (\texttt{2BBGABS} and \texttt{BBPLGABS}, respectively), also fit the data well ($\chi^2_\nu(\nu) = 1.02(21)$ and $\chi^2_\nu(\nu) = 0.87(21)$). However, for both models, the column density is a factor of 1.5--5 in excess of the Galactic value (considering the 1$\sigma$ confidence interval). For the \texttt{2BBGABS} case, the \texttt{BB} radius at 1~kpc distance is too large for an INS nature. Assuming a canonical radius of 12~km, this would imply distances between 40 and 400~pc that are in stark contrast to the high $\nh$ values. In the case of the \texttt{BBPLGABS} model, the photon index of the power-law component is unreasonably steep ($\Gamma = 12.4^{+1.6}_{-2.8}$), while the thermal component only dominates the high-energy end of the spectrum. This is at odds with what is observed in, X-ray binaries and rotation-powered pulsars, for instance, making these composite models overall less compelling than those with multiple lines.

The investigation of a hard excess in the spectrum of \otos\ is made difficult at the current signal-to-noise ratio by contamination from the neighbouring X-ray source. In order to constrain it, we extracted the EPIC-pn spectrum using a large extraction region radius of $\approx 40$\arcsec, which encompasses PSF counts from both sources. We then fitted the spectrum in the energy range of 0.2--8~keV. We defined a model consisting of an absorbed \texttt{BB} component with two absorption lines and an independently absorbed \texttt{PL} component, with both the column density and photon index parameters fixed to the best-fit values obtained for the contaminant. The fit is virtually identical to the \texttt{BB2GABS} fit in Table~\ref{tab_fitinfo}, with no indications of a systematic excess emission at harder energies. The \texttt{BB} and \texttt{PL} components converge to flux values of $4.62(5)\times 10^{-13}$~\fluxcgs\ and $8.8(6)\times 10^{-14}$~\fluxcgs\ (both 0.2--10~keV), respectively. These values agree well with those obtained from the single fits to \otos\ and the contaminant spectrum.

Using the period estimate from \xmm\ (Table~\ref{tab_timepar}), we performed phase-resolved spectroscopy with the EPIC-pn dataset by splitting the photons into four phase bins (0.875--0.125, 0.125--0.375, 0.375--0.625, and 0.625--0.875) including 2000--3000 photons each. The four spectra were then simultaneously fitted with a \texttt{BB}, \texttt{BBGABS}, and \texttt{BB2GABS} model. To lift the degeneracies and aid the fits, the $\nh$ was fixed to the phase-averaged upper limit of $5\times 10^{20}$~cm$^{-2}$, the line widths of the broad and narrow feature were fixed to $150$~eV and $15$~eV, respectively, and the effective temperature was tied to the same value for all phases. We found that a single \texttt{BB} model does not give a convincing fit ($\chi^2_\nu(\nu)=9.60(82)$), but the fit can be improved when broad lines at $\sim300$~eV are included at all phases ($\chi^2_\nu(\nu)=1.38(74)$). This fit shows strong residuals at $\sim 590$~eV that are most prevalent at phase 0.125--0.375. Including a second narrow line in this phase bin yields a near perfect fit ($\chi^2_\nu(\nu)=1.01(72)$; see Table~\ref{tab_phsresspec}). Fitting this narrow feature in other phase bins gives similar line energies around $590$~eV, but only marginally improves the fit statistic ($\chi^2_\nu\sim0.95$). We found that leaving the line widths or effective temperatures free to vary during the fits only slightly improves the fit statistic, but results in decreased precision in the parameter estimation, such that no significant changes in temperature or line width can be observed. We studied the detection significance of the features by simulating for each phase and feature 1000 spectra with and without the best-fit line. We then repeated the spectral analysis with the real and simulated data, including in each fit only one simulated spectrum at a time, to study the false-positive and false-negative rates. The line parameters of the broad features and the narrow feature at phase 0.125--0.375 were generally well recovered, and all fits implied low false-positive and false-negative rates virtually identical with zero. However, the current data do not allow us to significantly detect or exclude the existence of the narrow feature at the other phase bins because the simulations resulted in false-positive rates of 5--70~\% and false-negative rates of 8--47\%. We conclude that the broad feature appears to be observable in all phases. The narrow feature could be phase dependent because it was only significantly detected in one phase bin. However, additional observations are necessary to fully characterise its phase-dependence.

We tried simultaneous phase-averaged fits of all the available instruments (EPIC-pn, EPIC-MOS1, EPIC-MOS2, and \nicer) next. The results are generally consistent with the single EPIC-pn fits. The only difference is that thermal models, including only a single absorption line, already fit well (as shown for the \texttt{BB} case in Table~\ref{tab_fitinfo}). Systematic residuals, indicating the necessity for a second line around 600~eV, are not apparent. We attribute this to the fact that the noise added by including \nicer\ and EPIC-MOS dominates in comparison to the residuals caused by the absorption feature. Consequently, adding a second \texttt{GABS} component to the spectral fit improves the fit statistic only slightly and is less well constrained than EPIC-pn. We thus accept as our final model a pure blackbody continuum of $95.1^{+1.7}_{-2.6}$ eV modified by a broad line at $260^{+80}_{-50}$~eV, a narrow line at $587^{+8}_{-5}$~eV, and cold interstellar matter with a column density below $5\times 10^{20}$~cm$^{-2}$ (see line 3 in Table~\ref{tab_fitinfo}).


\begin{table*}
\caption{X-ray spectrum fitting results.
\label{tab_fitinfo}}
\centering
\scalebox{.86}{
\begin{tabular}{cccccccccccccc}
\hline\hline
\multicolumn{5}{l}{\texttt{BB}}\\
\hline
& $\nh$ & $kT$ & Radius\tablefootmark{(a)} & $\epsilon_1$ & $\sigma_1$ & EW$_1$\tablefootmark{(b)} & $\epsilon_2$ & $\sigma_2$ & EW$_2$\tablefootmark{(b)} & $\chi^2_\nu(\nu)$ & Absorbed flux\tablefootmark{(c)}\\
& $[10^{20}$~cm$^{-2} ] $ & [eV] & [km] & [eV] & [eV] & [eV] & [eV] & [eV] & [eV] & & [$10^{-13}$~\fluxcgs]\\
\hline
pn & $27.3^{+1.6}_{-1.5}$ & $86.7 \pm 1.6$ & $11.2^{+1.8}_{-1.1}$ & & & & & & & 4.09(26) & $4.63(5)$\\
pn & $12(5)$ & $88.5^{+2.8}_{-2.7}$ & $8.4^{+2.4}_{-1.4}$ & $360^{+60}_{-90}$ & $140^{+50}_{-40}$ & $280^{+210}_{-130}$ & & & & 1.69(23) & $4.78(5)$ \\
pn & $<5$ & $95.1^{+1.7}_{-2.6}$ & $5.1^{+1.2}_{-0.5}$ & $260^{+80}_{-50}$ & $145^{+18}_{-40}$ & $400^{+130}_{-160}$ & $587^{+8}_{-5}$ & $11^{+22}_{-5}$ & $25^{+150}_{-9}$ & 1.04(20) & $4.88(5)$\\
pn & $<2$ & $94.7^{+1.6}_{-2.0}$ & $5.2^{+0.7}_{-0.5}$ & $248^{+16}_{-7}$ & $150$ & $406^{+19}_{-26}$ & $588^{+7}_{-8}$ & $15$ & $27^{+4}_{-5}$ & 0.95(22) & $4.87(5)$ \\
All & $30.1^{+1.2}_{-0.6}$ & $83.2 \pm 1.1$ & $14.2^{+1.5}_{-1.1}$ & & & & & & & 1.34(910) & $4.66(3)$ \\
All & $3.5^{+2}_{-1.1}$ & $88.8^{+1.9}_{-2.0}$ & $7.6^{+1.1}_{-0.8}$ & $<260$ & $198^{+8}_{-23}$ & $430^{+70}_{-260}$ & & & & 1.07(907) & $5.14(4)$\\
All & $1.9^{+4}_{-1.6}$ & $88.6 \pm 2.0$ & $7.3^{+1.1}_{-0.9}$ & $<240$ & $192^{+8}_{-19}$ & $440^{+140}_{-300}$ & $633^{+11}_{-10}$ & $<43$ & $9.0^{+400}_{-1.2}$ & 1.06(904) & $5.06(4)$ \\
\hline
\multicolumn{5}{l}{\texttt{PL}}\\
\hline
 & $\nh$ & $\Gamma$ &  & $\epsilon$ & $\sigma$ & EW\tablefootmark{(b)} & & & & $\chi^2_\nu(\nu)$ & Absorbed flux\tablefootmark{(c)}\\
 & [$10^{20}$~cm$^{-2}$] & & & [eV] &[eV] & [eV]&  &  &  & &  [$10^{-13}$~\fluxcgs]\\
\hline
pn & $86^{+5}_{-4}$ & $10.630^{+0.04}_{-0.026}$ & & & & & & & & 17.27(26) & $4.42(5)$\\
pn & $26^{+6}_{-8}$ & $9.5(4)$ & & $330^{+50}_{-90}$ & $171^{+50}_{-23}$ & $530^{+210}_{-150}$ & & & & 0.86(23) & $4.77(5)$ \\
\hline
\multicolumn{5}{l}{\texttt{NSA}\tablefootmark{(d)}}\\
\hline
& $\nh$ & $kT$ & B & $\epsilon_1$ & $\sigma_1$ & EW$_1$\tablefootmark{(b)} & $\epsilon_2$ & $\sigma_2$ & EW$_2$\tablefootmark{(b)} & $\chi^2_\nu(\nu)$ & Absorbed flux\tablefootmark{(c)}\\
 & [$10^{20}$~cm$^{-2}$] & [eV] & [G] & [eV] & [eV] & [eV] & [eV] & [eV] & [eV] & & [$10^{-13}$~\fluxcgs]\\
\hline
pn & $37.7^{+0.6}_{-0.8}$    & $20.53(19)$ & $0$       & & & & & & & 5.52(26) & $4.58(5)$\\
pn & $13.3^{+0.8}_{-0.6}$    & $21.35^{+0.20}_{-0.3}$ & $0$       & $338^{+9}_{-7}$   & $146(5)$ & $361^{+14}_{-12}$ & & & & 2.18(23) & $4.77(5)$ \\
pn & $<1.7$ & $23.5^{+0.8}_{-0.8}$ & $0$     & $234^{+28}_{-40}$   & $156^{+13}_{-15}$ & $440^{+70}_{-90}$ & $585^{+7}_{-8}$ & $18^{+17}_{-16}$ & $31^{+60}_{-23}$ & 1.39(20) & $4.89(5)$\\

pn & $42.5^{+2.2}_{-1.8}$    & $28.9^{+1.0}_{-1.1}$ & $10^{12}$ & & & & & & & 7.05(26) & $4.57(5)$\\
pn & $16^{+5}_{-6}$          & $30.8^{+1.9}_{-1.4}$ & $10^{12}$ & $350^{+50}_{-80}$ & $151^{+40}_{-28}$ & $380^{+180}_{-140}$ & & & & 1.41(23) & $4.77(5)$ \\
pn & $<9$   & $35.8^{+1.4}_{-2.4}$ & $10^{12}$ & $<370$ & $173^{+13}_{-27}$ & $460^{+190}_{-400}$ & $590^{+8}_{-7}$ & $<50$ & $33^{+4}_{-28}$ & 0.77(20) & $4.88(5)$ \\

pn & $43.9^{+1.8}_{-1.9}$    & $31.2^{+0.9}_{-0.8}$ & $10^{13}$ & & & & & & & 7.07(26) & $4.56(5)$\\
pn & $15(5)$          & $33.5^{+1.4}_{-1.0}$ & $10^{13}$ & $350^{+50}_{-70}$ & $154^{+40}_{-29}$ & $400^{+180}_{-160}$ & & & &1.30(23) & $4.78(5)$ \\
pn & $<9$   & $38.4^{+1.1}_{-5}$  & $10^{13}$ & $<350$ & $172^{+16}_{-80}$ & $470^{+170}_{-400}$ & $591(8)$ & $<50$ & $36^{+14}_{-28}$ & 0.71(20) & $4.89(5)$ \\
\hline
\multicolumn{5}{l}{\texttt{2BBGABS}}\\
\hline
& $\nh$ & $kT_1$ & Radius$_1$\tablefootmark{(a)} & $kT_2$ & Radius$_2$\tablefootmark{(a)} & & $\epsilon$ & $\sigma$ & EW\tablefootmark{(b)} & $\chi^2_\nu(\nu)$ & Absorbed flux\tablefootmark{(c)}\\
& $[10^{20}$~cm$^{-2} ] $ & [eV] & [km] & [eV] & [km] &  & [eV] & [eV] & [eV] & & [$10^{-13}$~\fluxcgs]\\
\hline
pn & $15^{+6}_{-5}$ & $60^{+16}_{-10}$ & $80^{+240}_{-50}$ & $135^{+50}_{-18}$ & $0.8^{+1.4}_{-0.4}$ & & $<400$ & $230^{+170}_{-80}$ & $600^{+600}_{-500}$ & 1.02(21) & $4.77(5)$ \\

\hline
\multicolumn{5}{l}{\texttt{BBPLGABS}}\\
\hline
& $\nh$ & $kT$ & Radius\tablefootmark{(a)} & $\Gamma$ & & & $\epsilon$ & $\sigma$ & EW\tablefootmark{(b)} & $\chi^2_\nu(\nu)$ & Absorbed flux\tablefootmark{(c)}\\
& $[10^{20}$~cm$^{-2} ] $ & [eV] & [km] &  &  & & [eV] & [eV] & [eV] & & [$10^{-13}$~\fluxcgs]\\
\hline
pn & $23^{+8}_{-5}$ & $140^{+70}_{-50}$ & $0.5^{+4}_{-0.4}$ & $12.4^{+1.6}_{-2.8}$ & & & $<330$ & $239^{+18}_{-80}$ & $660^{+190}_{-500}$ & 0.87(21) & $4.78(5)$ \\
\hline

\end{tabular}
}
\tablefoot{We give the $1\sigma$ confidence intervals for the estimated parameters.
\tablefoottext{a}{We assumed a 1~kpc distance for the blackbody emission radius at infinity.}
\tablefoottext{b}{The equivalent width (EW) was estimated via $\int \frac{F_c-F_o}{F_c} dE$, with $F_c$ being the continuum and $F_o$ the observed flux. The errors state the maximum and minimum EW values obtained from all possible combinations of the upper and lower $1\sigma$ confidence interval limits of the model parameters.}
\tablefoottext{c}{The absorbed model flux covers the 0.2--10~keV range.}
\tablefoottext{d}{The model parameters assume a canonical neutron star with 1.4~M$_\odot$ and 12~km radius.}
}
\end{table*}



\begin{table}
\caption{Phase-resolved spectrum fit result.
\label{tab_phsresspec}}
\centering
\scalebox{.94}{
\begin{tabular}{cccccccccccccc}
\hline\hline
Phase & Radius\tablefootmark{(a)} & $\epsilon_1$ & EW$_1$\tablefootmark{(b)} & $\epsilon_2$ & EW$_2$\tablefootmark{(b)}\\
 & [km] & [eV] & [eV] & [eV] & [eV]\\
\hline
0.875--0.125 & $5.1^{+0.5}_{-0.4}$ & $278^{+25}_{-30}$ & $330^{+50}_{-50}$ & & \\
0.125--0.375 & $7.4^{+0.7}_{-0.6}$ & $273^{+23}_{-27}$ & $410^{+50}_{-50}$ & $594^{+9}_{-9}$ & $41^{+7}_{-7}$ \\
0.375--0.625 & $6.2^{+0.6}_{-0.5}$ & $315^{+16}_{-19}$ & $356^{+26}_{-26}$ & & \\
0.625--0.875 & $6.6^{+0.7}_{-0.5}$ & $305^{+16}_{-18}$ & $366^{+27}_{-25}$ & & \\
\hline
\end{tabular}
}
\tablefoot{We give the $1\sigma$ confidence intervals for the estimated parameters. The $\nh$ is set to $5\times 10^{20}$~cm$^{-2}$, the line widths are set to $150$~eV and $15$~eV for the first and second line, respectively. The fit converged to a temperature of $kT=91.8^{+1.5}_{-1.5}$~eV and fit statistic of $\chi^2_\nu(\nu)=1.01(72)$.
\tablefoottext{a}{We assumed a 1~kpc distance for the blackbody emission radius at infinity.}
\tablefoottext{b}{The equivalent width (EW) was estimated the same way as for the phase-averaged fits. See notes of Table~\ref{tab_fitinfo} for more details.}
}
\end{table}


\subsection{VLT observation}
\begin{figure}[t]
\begin{center}
\includegraphics[width=\linewidth]{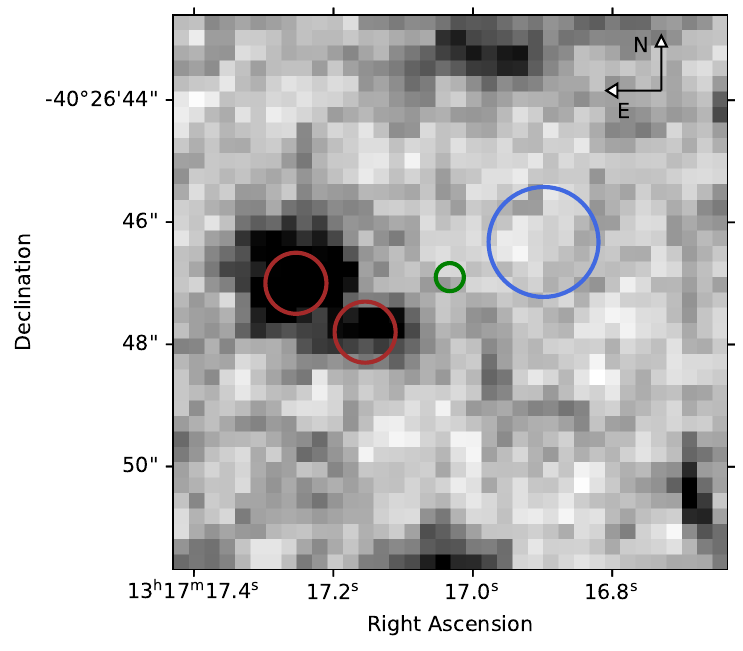}\vskip1pt
\end{center}
\caption{FORS2 \textit{R\_SPECIAL}-band image of the field of \otos. The circles indicate the 1$\sigma$ positional uncertainty estimated from \xmm\ (green) and \eros\ (blue). We searched for possible counterparts using the \texttt{SExtractor} software. The identified field objects are marked in brown (circles have arbitrary radii of 1\arcsec).}
\label{fig_vlt_img}
\end{figure}

With respect to previous imaging with the \saltl\ \citep[SALT;][]{2023A&A...674A.155K}, the deeper \eso\ observations allow us to further improve the limit on the X-ray-to-optical flux ratio for this source and to search for faint optical counterparts. We applied the \texttt{SExtractor} software \citep{1996A&AS..117..393B} to detect all sources near \otos\ and mark them in Fig.~\ref{fig_vlt_img}. The updated \xmm\ EPIC position of the target, although consistent within 2$\sigma$, is 1.3\arcsec\ offset from that derived from the \eros\ data. This brings \otos\ closer to two field sources that are now separated by only 1.7\arcsec\ and 2.5\arcsec\ (see sources to the east in Fig.~\ref{fig_vlt_img}). The magnitudes of these sources were measured at 26~mag and 24.5~mag, respectively. They were not detected in the SALT image presented in \citet[][]{2023A&A...674A.155K}, since their magnitudes are close to the detection limit. Nonetheless, the association of the target with any of these possible optical counterparts is unlikely, with angular separations at $7\sigma$ and $11\sigma$ (Table~\ref{tab_posphotpar}).

\begin{figure}[t]
\begin{center}
\includegraphics[width=\linewidth]{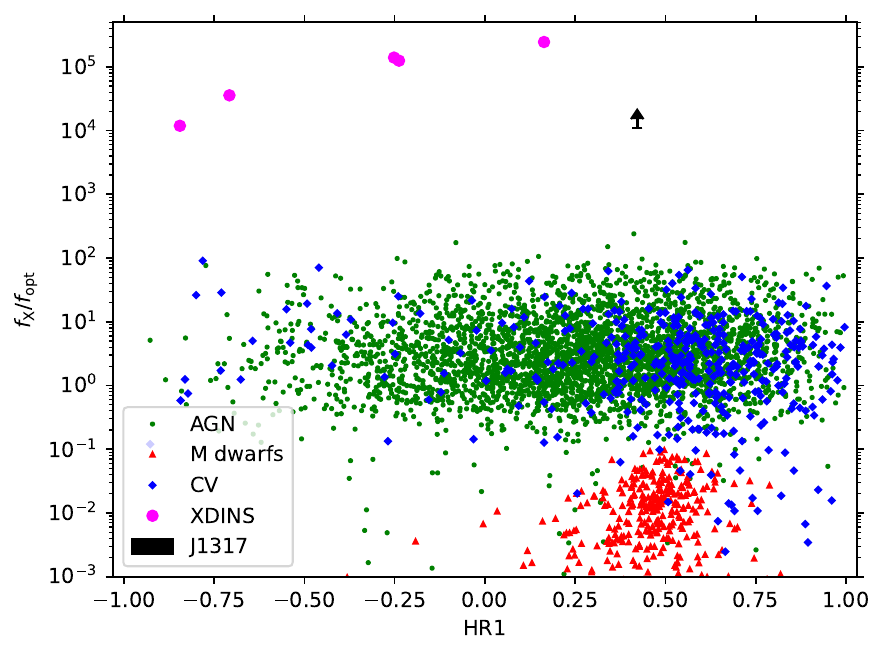}\vskip1pt
\end{center}
\caption{X-ray-to-optical flux ratios of several classes of X-ray emitters as a function of hardness ratio \citep[see][for details]{2023A&A...674A.155K}. The updated position of the target is denoted by the lower limit in black and is compared to those of the five XDINSs on the German eROSITA footprint (western hemisphere from the Galactic centre; magenta data points).}
\label{fig_spec_plots}
\end{figure}

We used the nightly determined calibration zero points and extinction values provided by ESO\footnote{The magnitude zero points and extinction values can be accessed at: \url{http://archive.eso.org/qc1/qc1_cgi?action=qc1_browse_table&table=fors2_photometry}} for the magnitude calibration of the FORS imaging. Assuming a $5\sigma$ detection limit, the equation for the magnitude limit of a point source in an optimal Gaussian aperture was used to estimate the image depth. Based on the parameters shown in Table~\ref{tab_posphotpar}, we obtained a limiting magnitude of 27.5~mag at the position of the target, resulting in an X-ray-to-optical flux ratio of $10^4$ at least. Such an extreme value can only be observed among INSs (Fig.~\ref{fig_spec_plots}).

\section{Discussion and outlook\label{sec_disc}}

In this work, we report the results of dedicated \nicer, \xmm, and \eso\ follow-up observations of the INS candidate \otos. This campaign unveiled a highly significant periodic signal at a period of $12.757$~s, very likely associated with the spin of the neutron star. The pulse profile is clearly double-humped, and we found the pulsed fraction to increase with energy. The results are well in line with expectations for a highly magnetised INS.

\begin{figure}[t]
\begin{center}
\includegraphics[width=\linewidth]{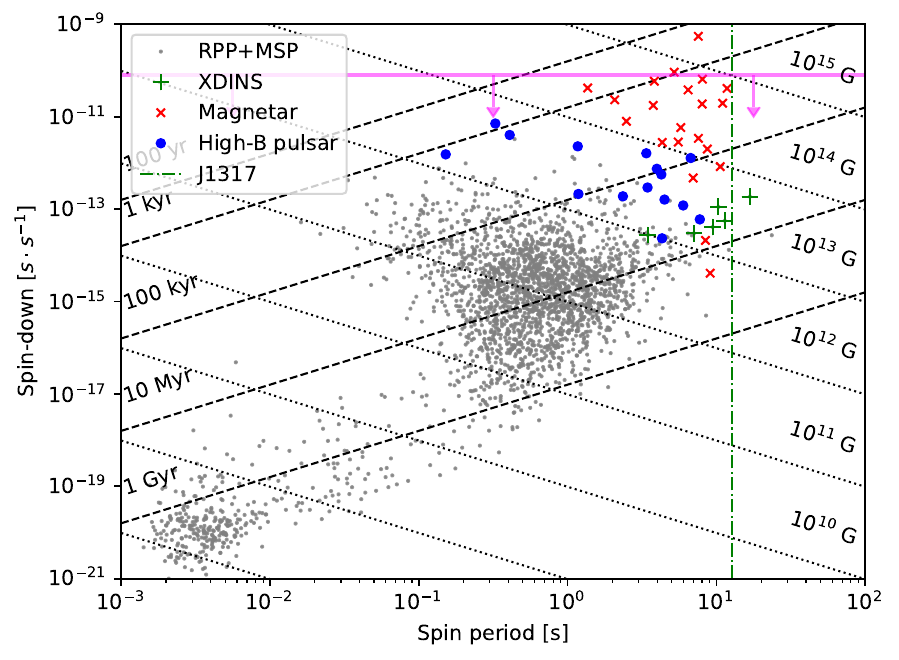}\vskip1pt
\end{center}
\caption{Spin vs. spin-down diagram presenting the spin properties of the known isolated neutron star population. In the background, we plot all the objects collected in the ATNF pulsar database \citep[Version 1.70;][]{2005AJ....129.1993M}. The period of \otos\ is marked by the vertical green line. The horizontal magenta line indicates the upper 3$\sigma$ spin-down limit derived from the \xmm\ and \nicer\ observations.}
\label{fig_p_pdot}
\end{figure}

In Fig.~\ref{fig_p_pdot} we plot the spin versus spin-down diagram of neutron stars from the ATNF Pulsar Catalogue\footnote{\url{https://www.atnf.csiro.au/people/pulsar/psrcat}} \citep[Version 1.70;][]{2005AJ....129.1993M}. The detected periodicity places \otos\ where we would expect to find an XDINS, magnetar or high-B pulsar \citep[see e.g.][and references therein, for an overview of the Galactic INS population]{2019RPPh...82j6901E}. The \xmm\ and \nicer\ epochs only allow us to derive an upper limit on the spin-down of the source, $\dot{P} \leq 8\times 10^{-11}$~s~s$^{-1}$ ($3\sigma$). Under the usual assumption of magnetic dipole braking in vacuum \citep{1969ApJ...157.1395O}, the dipolar magnetic field in the equator and characteristic age of \otos\ are constrained to $B_\mathrm{dip} = 3.2\times10^{19}\sqrt{P\dot{P}}$~G~$\leq 10^{15}$~G and $\tau_\mathrm{ch} \geq 2600$~yr. A better constraint on the evolutionary state of the neutron star relies on the eventual measurement of the spin-down rate and possible detection of radio and $\gamma$-ray counterparts.

The deep \eso\ FORS2 observations of the field of the XDINS candidate in the \textit{R\_SPECIAL} band revealed two faint sources in the vicinity of \otos. The association of these two sources with the neutron star candidate is unlikely however, because the updated \xmm\ EPIC position implies a significant separation of 7--11$\sigma$ from these objects. The absence of counterparts to such an extreme X-ray-to-optical flux ratio, $\geq 10^4$, further supports an INS nature (Fig.~\ref{fig_spec_plots}).

The spectrum of \otos\ cannot be described by a single-component model. Good spectral fits can be obtained if the continuum (a blackbody, power-law, or a fully ionised neutron star model atmosphere) is modified by one or two Gaussian absorption features. All these models have in common that one line is located at soft energies between 200~eV and 400~eV with a width of $\sigma \sim 150$~eV. The second line, strictly necessary for thermal continuum fits (\texttt{BB} and \texttt{NSA}) to the EPIC-pn spectrum, is located around $590$~eV with a smaller Gaussian width of $\sigma = 10-30$~eV. The lines seem to be detected with very high significance. Testing other absorption components, we found that only an \texttt{edge} model affects the spectral results significantly because a third feature at $\sim860$~eV is necessary to achieve an acceptable fit. However, we note that compared to the models using a \texttt{GABS} component, the fit statistic is generally worse and the $\nh$ significantly exceeds the Galactic value, making them less convincing. The temperature of the continuum spectrum is $kT \sim 95$~eV in the \texttt{BB} case and $kT\sim 35$~eV for the \texttt{NSA} model. We note that the low \texttt{NSA} model temperature may predict an optical flux that is too high, as is known from the XDINSs \citep{1996ApJ...472L..33P}. Similarly, the \texttt{NSA} model assumes uniform temperature emission from the full neutron star surface, which might explain the low distance values from this model. Other spectral models may give statistically valid fits, but are less compelling on the basis of their parameter values (e.g. a \texttt{2BBGABS} model that implies a small distance of only 40--400~pc for a typical neutron star radius of 12~km, but, despite the absorption feature, converges to column densities well in excess of the Galactic value).

The \eros\ spectra discussed in \cite{2023A&A...674A.155K} did not allow us to statistically exclude a \texttt{PL} nature of the source. Here, a good fit can be obtained when a \texttt{PL} is combined with an absorption feature, but the resulting photon index of $\Gamma = 9.5(4) $ is even larger than observed for \eros. This value is too high because neutron star non-thermal emission components are usually observed to have $\Gamma <3$ \citep[][]{2009ASSL..357...91B}. The column density exceeding the Galactic value may imply an extragalactic nature. The only extragalactic objects that reach very large photon indices are Seyfert I galaxies \citep{2022arXiv221106184G}. However, the very high X-ray-to-optical flux limits and the strong detected pulsation at 12.757~s make an Seyfert I galaxy nature very unlikely. Thus, it seems that the \texttt{PL} can be rejected as a valid spectral model.

The obtained flux values from the EPIC-pn spectral fits and from fits using all spectra (Table~\ref{tab_fitinfo}) agree with those derived from \eros\ \citep{2023A&A...674A.155K} and thus by extension also with \ros. Nevertheless, if the long-term variability of \otos\ is to be probed, the influence of the contaminant needs to be considered, which was not separately detected in \ros, in any of the \eros\ surveys, or in the independent analysis of \citet{2023A&A...674A.155K}. The best-fit contaminant power-law model indicates a flux of $\sim 3\times10^{-14}$~\fluxcgs\ in the 0.2--2~keV band, where the emission is most prevalent in the short \eros\ observation. This is of the same size as the $1\sigma$ \eros\ flux error. Since the \eros\ confidence interval agrees well with results from \xmm\ and \nicer, we conclude that the contaminant did not significantly affect the \eros\ flux measurement. Since the estimated \eros\ count rate is comparable with the \ros\ measurement (with a similarly short exposure), the emission of \otos\ seems to have been stable over the past 30 years.

The \texttt{BB} fits give upper limits on the absorbing hydrogen column density and can therefore be used to compute limits on the distance to the source. We used the relations given in \citet{2009MNRAS.400.2050G} and \citet{1989ApJ...345..245C} to compute the total extinction $A(5500\mathrm{\AA}) < 0.226$ for $\nh < 5\times 10^{20}$~cm$^{-2}$. We compared this limit to the total extinction in the direction of \otos\ as it can be inferred from the GAIA-2MASS 3D maps of Galactic interstellar dust \citep{2022A&A...661A.147L} and found them to be almost exactly in accordance with a distance of $1$~kpc. This implies a radius of $5.1^{+1.2}_{-0.5}$~km and a luminosity of $2.7^{+1.3}_{-0.6} \times 10^{32}$~erg~s$^{-1}$. More constraining values can be derived from the spectral fit with fixed Gaussian line widths. Here, the limit of $\nh<2\times 10^{20}$~cm$^{-2}$ implies $A(5500\mathrm{\AA}) < 0.09$ and indicates a distance of only 185~pc. We then computed a \texttt{BB} radius of $0.96^{+0.13}_{-0.09}$~km and a luminosity of $9.6^{+2.5}_{-1.8} \times 10^{30}$~erg~s$^{-1}$. We note that similarly to the \texttt{NSA} distance estimate, the distance of $185$~pc may be too small. While the value would agree with the distance to the closest XDINS \magoe\ \citep{2007Ap&SS.308..191V}, we note that the flux of \otos\ is a factor 4--20 lower than those observed for the other XDINSs ($\sim 2\times10^{-12}-1\times10^{-11}$~\fluxcgs). Only the flux of the XDINS \magzf\ may be comparable with \otos\ \citep{2004A&A...424..635H}, but the spectrum of this source is much softer, and its distance was estimated to $\sim345$~pc \citep{2007Ap&SS.308..171P}.

The fact that the thermal continuum needs to be modified by multiple absorption features to properly fit the spectrum makes \otos\ very similar to XDINSs and other thermally emitting INSs. In particular, a very similar broad absorption line at $200(40)$~eV with a Gaussian width of $\sigma = 139^{+12}_{-13}$~eV and equivalent width (EW) of $179^{+3}_{-59}$~eV has been reported in the EPIC-pn spectrum of \magot\ \citep{2017MNRAS.468.2975B}. Interestingly, \magot\ also exhibits a double-humped pulse profile at a similar long spin period of 10.31~s, and it is the source with the highest pulsed fraction of the seven XDINSs ($19$\%, 0.2--1.2~keV). As observed for \otos, the pulsed fraction seems to increase with energy \citep{2005A&A...441..597S}. The other XDINSs possess features with lower EW values that agree better with the more narrow line at $\sim 590$~eV \citep[e.g.][]{2009ASSL..357..141T}.

Absorption features in the spectra of XDINSs are expected to arise due to proton cyclotron resonances \citep{2019A&A...622A..61S}, atomic bound-bound or bound-free transitions in the neutron star atmosphere \citep[e.g.][]{2007Ap&SS.308..191V}, the presence of highly ionised oxygen in the ISM, atmosphere, or vicinity of the neutron star \citep{2009A&A...497L...9H, 2012MNRAS.419.1525H}, or they might be of spurious nature because multi-temperature thermal components on the surface of the neutron star might mimic such features \citep{2014MNRAS.443...31V}. The general consensus is that the spectra combine features with different physical origins, and thus, the two detected features in the spectrum of \otos\ might generally be of different nature. When we interpret them to be caused by proton cyclotron absorption, the magnetic field of \otos\ could be estimated via $E_{cyc} = 0.063 \frac{B}{10^{13}\mathrm{G}} \times(1+z) \left[\mathrm{keV}\right]$, with $z$ being the gravitational redshift. For a canonical neutron star with $M=1.4$~M$_\odot$ and $R=12$~km, $z = 0.35$. Based on the line energy of $260^{+80}_{-50}$~eV, a proton cyclotron feature would imply a magnetic field of $5.6^{+1.8}_{-1.1} \times 10^{13}$~G. This would indicate that \otos\ is a highly magnetised neutron star, and it agrees well with the dipolar magnetic field strengths inferred for XDINSs. The line energy of the second feature would imply a magnetic field strength of $1.255^{+0.018}_{-0.011} \times 10^{14}$~G, indicating that it could emerge from a more magnetised structure near the neutron star surface. This configuration was proposed for narrow absorption features in other thermally emitting INSs \citep[e.g.][]{2013Natur.500..312T, 2017MNRAS.468.2975B}. However, these features are usually too weak to be detectable in the phase-averaged spectrum and were observed to be highly phase dependent. Given the results of the phase-resolved spectroscopy, establishing or excluding a phase-dependent nature of the narrow feature of \otos\ is difficult, however. Alternatively, the features of \otos\ might be interpreted as being produced by atomic transitions in a hydrogen atmosphere, but in this case, the feature at $\sim 590$~eV would also need magnetic field strengths in excess of $10^{14}$~G \cite[see Fig.~7 in][]{2007Ap&SS.308..191V}, which is higher than the dipolar fields observed in the other XDINSs. Thus, the presence of a He or higher-Z atmosphere would be more compelling because they allow for lines at the observed energy and typical XDINS magnetic field strengths \citep[see e.g.][]{2005ApJ...635L..61P, 2007MNRAS.377..905M}. Absorption by the presence of highly ionised oxygen would cause features between $500-600$~eV. These lines are generally narrower than the line that is observed for \otos, which might imply that it is intrinsic to the source \citep[comparable to the interpretation of the line at $\sim530$~eV in the RGS spectrum of \magot;][]{2012MNRAS.419.1525H}. In the last case, the features could be artefacts caused by a non-optimal spectral modelling that does not sufficiently account for the effects of a more complicated surface temperature distribution. However, \cite{2014MNRAS.443...31V} reported that the spectrum of \magot\ is difficult to model with inhomogeneous temperature distributions. Because the spectra of \otos\ and \magot\ are very similar, this might also hold true here. Based on the available data alone, an exact characterisation of the observed features is difficult because at the moment, multiple production mechanisms are possible for each of them. Additional observations are required to attain more stringent constraints on their parameters (e.g. phase dependence, energy, width, and strength).

\begin{figure}[t]
\begin{center}
\includegraphics[width=\linewidth]{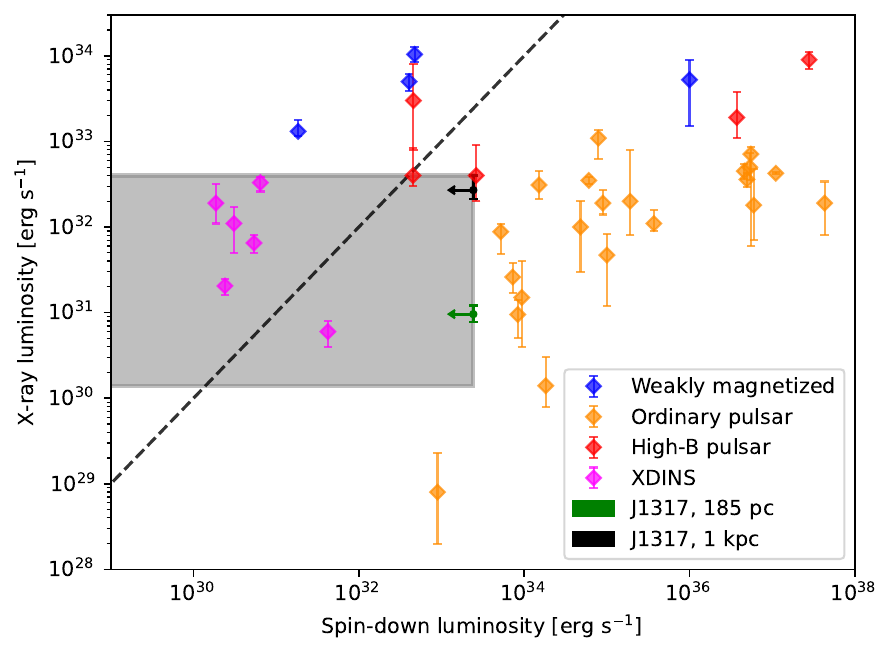}\vskip1pt
\end{center}
\caption{X-ray luminosity as a function of spin-down luminosity for the collection of thermally emitting isolated neutron stars from \citet{2020MNRAS.496.5052P}. The luminosity limits inferred for \otos\ are shown in green (for a distance of 185~pc) and black (for a distance of 1~kpc). The grey box indicates the parameter space constrained by the current data, assuming distances within 0.1--1~kpc and the spin-down limit from the timing analysis $\dot{P}\leq8\times10^{-11}$~s~s$^{-1} (3\sigma)$. The dashed black line indicates where both luminosities are equal in value.}
\label{fig_lum_comp}
\end{figure}


The soft and seemingly purely thermal spectrum and lack of significant flux and spectral variability speak against a magnetar nature. These sources, usually discovered when undergoing bright bursts of X-ray and soft $\gamma$-ray emission, possess much harder spectra, show higher persistent X-ray fluxes and strong and complex variability. Similarly, the absence of a detectable supernova remnant at X-ray energies seems to exclude a young pulsar nature. The XDINS population is oftentimes compared to the population of high-B pulsars \citep[see][and references therein]{2019RPPh...82j6901E}. Located at the intersection of conventional rotation-powered pulsars and magnetars (we indicated high-B pulsars mentioned in \citet{2011AIPC.1379...60N}, \citet{2013ApJ...764....1O}, and \citet{2020MNRAS.496.5052P} in Fig.~\ref{fig_p_pdot}), these objects possess magnetic fields of $B_\mathrm{dip} = 10^{13}-10^{14}$~G. However, their emission is thought to be governed by magnetic braking and not by remnant heat and the decay of the magnetic field, as is proposed for the XDINSs. In comparison with the latter, high-B pulsars seem less efficient in converting spin-down power into X-ray luminosity, as is the case for conventional pulsars \citep[e.g.][]{2002A&A...387..993P}. Based on the spin-down rate limit derived for \otos\ (Sect.~\ref{sec_timing}), we can constrain its spin-down luminosity as $\dot{E} ~\sim 6.3\times 10^{46} (\dot{P}P^{-3})$~g~cm$^{2} \leq 2.4 \times 10^{33}$~erg~s$^{-1}$ at the $3\sigma$ level. For comparison, in Fig.~\ref{fig_lum_comp} we show the X-ray luminosity of several isolated pulsars as a function of the luminosity available from spin down. The parameter space covered by our current limits on \otos\ is marked by a shaded grey area, assuming that the source is at a distance of 0.1--1~kpc. Based on the current limits, both a high-B pulsar and an XDINS are possible for \otos. This makes the neutron star an interesting target for a search for its radio counterpart. On the one hand, the detection of coherent radio emission may provide an independent distance estimate from the dispersion measure and a possible constraint on the pulsar braking index via a precise timing solution. On the other, it can shed light on whether XDINSs are truly radio quiet or are nearby long-period radio pulsars for which the narrow emission beam simply does not sweep over the Earth \citep{2009ApJ...702..692K}. In comparison to the known high-B pulsars ($0.15$~s~$\leq P\leq 7.74$~s), the spin period of \otos\ is longer. At the time of writing, \otos\ is only the 13$^\mathrm{th}$ neutron star detected with a spin period in excess of 10~s. Only the XDINS \magzs\ \citep{2017A&A...601A.108H} and the radio detected neutron stars \gleam\ \citep{2022Natur.601..526H}, \onoo\ \citep{2022NatAs...6..828C}, \osts\ \citep{2018ApJ...866...54T}, and \onot\ \citep{2021RAA....21..107H} possess longer spin periods. Combined with its bright thermal luminosity, which is already conservatively comparable to that available from spin-down, this suggests that the evolutionary path is considerably affected by the super strong magnetic field of the neutron star.

To summarise, the recent observations of \otos\ have clearly confirmed the INS nature of the source. The detected pulsations at 12.757~s and evidence of spectral absorption features are remarkable among X-ray pulsars and may indicate the signature of a strong magnetic field. Its overall properties are in line with an XDINS nature within the scenario that they have evolved from magnetars, but they are also similar to those of high-B pulsars for which the effects of field decay have not been as significant. The \xmm\ and \nicer\ observations give only a weak limit on the spin-down rate of the neutron star. A dedicated coherent timing solution is necessary to further characterise the evolutionary state of the source, that is, to determine its position in the spin versus spin-down diagram of INSs and obtain more stringent estimates on the neutron star magnetic field, characteristic age, and spin-down luminosity. Similarly, observations at radio energies might be interesting because the apparent lack of radio emission in XDINSs is one of their main defining characteristics.

\begin{acknowledgements}

We thank the anonymous referee and the editor for very helpful comments and suggestions that improved this article.

This work was funded by the Deutsche Forschungsgemeinschaft (DFG, German Research Foundation) – 414059771. This work was supported by the project XMM2ATHENA, which has received funding from the European Union's Horizon 2020 research and innovation programme under grant agreement n$^{\rm o}101004168$.

Based on observations made with ESO Telescopes at the La Silla Paranal Observatory under programme ID 111.259R.001

This research has made use of data and/or software provided by the High Energy Astrophysics Science Archive Research Center (HEASARC), which is a service of the Astrophysics Science Division at NASA/GSFC.

This work is based on data from eROSITA, the soft X-ray instrument aboard SRG, a joint Russian-German science mission supported by the Russian Space Agency (Roskosmos), in the interests of the Russian Academy of Sciences represented by its Space Research Institute (IKI), and the Deutsches Zentrum für Luft- und Raumfahrt (DLR). The SRG spacecraft was built by Lavochkin Association (NPOL) and its subcontractors, and is operated by NPOL with support from the Max Planck Institute for Extraterrestrial Physics (MPE).

The development and construction of the eROSITA X-ray instrument was led by MPE, with contributions from the Dr. Karl Remeis Observatory Bamberg \& ECAP (FAU Erlangen-Nuernberg), the University of Hamburg Observatory, the Leibniz Institute for Astrophysics Potsdam (AIP), and the Institute for Astronomy and Astrophysics of the University of Tübingen, with the support of DLR and the Max Planck Society. The Argelander Institute for Astronomy of the University of Bonn and the Ludwig Maximilians Universität Munich also participated in the science preparation for eROSITA.

The eROSITA data shown here were processed using the eSASS/NRTA software system developed by the German eROSITA consortium.

For analysing X-ray spectra, we use the analysis software BXA \citep{2014A&A...564A.125B}, which connects the nested sampling algorithm UltraNest \citep{2021JOSS....6.3001B} with the fitting environment XSPEC \citep{1996ASPC..101...17A}.

This research has made use of the VizieR catalogue access tool, CDS, Strasbourg, France (DOI : 10.26093/cds/vizier). The original description of the VizieR service was published in 2000, A\&AS 143, 23.

This research has used data, tools or materials developed as part of the EXPLORE project that has received funding from the European Union’s Horizon 2020 research and innovation programme under grant agreement No 101004214.

This work made use of Astropy:\footnote{http://www.astropy.org} a community-developed core Python package and an ecosystem of tools and resources for astronomy \citep{astropy:2013, astropy:2018, astropy:2022}.

\end{acknowledgements}
\bibliographystyle{aa}
\bibliography{ref_j1317}

\begin{appendix}
\end{appendix}

\typeout{get arXiv to do 4 passes: Label(s) may have changed. Rerun}

\end{document}